\documentclass[5p,times]{elsarticle}

\usepackage{amssymb}
\usepackage{graphicx}
\usepackage{makecell}
\usepackage{subcaption}
\usepackage{array}
\usepackage{multirow}
\usepackage{algorithm}
\usepackage{algpseudocode}
\usepackage{amsmath}
\usepackage{booktabs}
\usepackage{tablefootnote}
\usepackage{listings}
\usepackage[switch]{lineno}

\journal{Elsevier}

\begin{document}

\begin{frontmatter}



\title{XGV-BERT: Leveraging Contextualized Language Model and Graph Neural Network for Efficient Software Vulnerability Detection}

\author[inst1,inst2]{Vu Le Anh Quan}\ead{19520233@uit.edu.vn}
\author[inst1,inst2]{Chau Thuan Phat}\ead{19520827@gm.uit.edu.vn}
\author[inst1,inst2]{Kiet Van Nguyen}\ead{kietnv@uit.edu.vn}
\author[inst1,inst2]{Phan The Duy}\ead{duypt@uit.edu.vn}
\author[inst1,inst2]{Van-Hau Pham}\ead{haupv@uit.edu.vn}

\affiliation[inst1]{organization={Information Security Laboratory, University of Information Technology},
            city={Ho Chi Minh city},
            country={Vietnam}}

\affiliation[inst2]{organization={Vietnam National University Ho Chi Minh City},
            city={Hochiminh City},
            country={Vietnam}}

\begin{abstract}
With the advancement of deep learning (DL) in various fields, there are many attempts to reveal software vulnerabilities by data-driven approach. Nonetheless, such existing works lack the effective representation that can retain the non-sequential semantic characteristics and contextual relationship of source code attributes. Hence, in this work, we propose XGV-BERT, a framework that combines the pre-trained CodeBERT model and Graph Neural Network (GCN) to detect software vulnerabilities. By jointly training the CodeBERT and GCN modules within XGV-BERT, the proposed model leverages the advantages of large-scale pre-training, harnessing vast raw data, and transfer learning by learning representations for training data through graph convolution. The research results demonstrate that the XGV-BERT method significantly improves vulnerability detection accuracy compared to two existing methods such as VulDeePecker and SySeVR. For the VulDeePecker dataset, XGV-BERT achieves an impressive F1-score of 97.5\%, significantly outperforming VulDeePecker, which achieved an F1-score of 78.3\%. Again, with the SySeVR dataset, XGV-BERT achieves an F1-score of 95.5\%, surpassing the results of SySeVR with an F1-score of 83.5\%.
\end{abstract}




\end{frontmatter}


\section{Introduction}
Recently, as technology continues its rapid evolution, the software development landscape has witnessed an exponential surge. While these software innovations offer unprecedented convenience, they also bring forth a looming specter: a problem of software vulnerabilities. These vulnerabilities are formidable adversaries to the seamless functioning of software systems \cite{zhu2023application_SoftSec_Review}. The global economic toll, both direct and indirect, inflicted by these vulnerabilities has surpassed billions of dollars, making it an issue of paramount concern. It is an undeniable fact that the vast majority of software applications harbor various types of vulnerabilities. Notable among these are buffer overflow vulnerabilities like CVE-2019-8917, library/API function call vulnerabilities like CVE-2016-10033, array usage vulnerabilities like CVE-2020-12345, and many more extensively cataloged within the Common Vulnerabilities and Exposures (CVE) database \cite{cve}. As time elapses, the longer a vulnerability persists unaddressed, the more it becomes an inviting target for malicious actors. This, in turn, exposes companies and organizations to the ominous specter of substantial damages \cite{yonghee2010}. Consequently, the quest for the automated detection of software vulnerabilities within stringent timeframes stands at the vanguard of advanced research endeavors \cite{zeng2020software_DL_sur}, \cite{hanif2021riseML_soft}, \cite{le2022survey}.

On the other hand, Deep Learning (DL) technology can provide the capability to achieve more accurate automatic vulnerability detection \cite{marjanov2022machine}, \cite{lin2020software}, \cite{khan2021systematic}, \cite{zou2020muy}. With the continuous innovation and development of DL technology, significant advancements have been made in Natural Language Processing (NLP). Models such as GPT \cite{alec2018pretrain} and BERT \cite{jacob2018bert} have propelled NLP technology forward. Source code is essentially a text in a specific format, making it logically feasible to utilize NLP techniques for code analysis. In fact, models like CodeBERT \cite{zhangyin2020codebert} have been proposed by several researchers, and some code-level tasks have been addressed, yielding promising results. These findings demonstrate the potential of using NLP technology for automated vulnerability detection research.

However, there are various directions of research that employ DL for security vulnerability detection. According to the survey by Zeng et al. \cite{peng2020survey}, there are four main research directions. The first involves using DL models to learn the semantic representations of programs, as proposed by Wang et al. \cite{wang2016semantic}. The second direction focuses on end-to-end solutions for detecting Buffer Overflow vulnerabilities, as explored by Choi et al. \cite{choi2017buffer}. The third direction involves extracting vulnerability-containing code patterns to train models, as demonstrated by Li et al. \cite{li2018vuldeepecker}. Finally, the fourth direction addresses vulnerability detection for binary code, as studied by Liu et al. \cite{liu2020binary}.

Each of these research directions has its own advantages and limitations. Based on the previous research outcomes, extracting vulnerability-containing patterns to create a training dataset has achieved promising results, displaying relatively good effectiveness and potential for further development \cite{croft2022data_softsec_data_sur}. Notable examples of this approach include the VulDeePecker paper by Li et al. \cite{li2018vuldeepecker} and SySeVR by Li et al. \cite{li2021sysevr}, as well as VulDeBERT by Kim et al. \cite{kim2022vuldebert}. However, both SySeVR and VulDeBERT still exhibit certain shortcomings; namely, the processed data consists solely of isolated code statements extracted from the source code, lacking contextual linkage. This deficiency inherently diminishes the precision of the model. 

Meanwhile, to retain the non-sequential semantic attributes inherent in the source code of the program, certain graph-based methodologies have been introduced, as documented by \cite{wang2021_graph_vul}, \cite{Cao2022_MVD_graph}, \cite{wang2023_GraphSPD}, \cite{Wu2022_VulCNN}, \cite{hin2022linevd}, \cite{nguyen2022_ReGVD_graph}, \cite{guo2022hyvuldect}, \cite{tang2023csgvd_graph}. These studies advocate the transformation of contract source code into a graph representation, followed by the utilization of graph neural networks for vulnerability detection. These existing works indicate that there is the potential for utilizing graph representations for inspecting the relationship of source code components to reveal software vulnerability. Specifically, the slices extracted from the source code are transformed into graph representations and subsequently incorporated into the deep learning model for training. The conversion of slices into graphs aims to capture relationships between words, as well as between words and slices, enhancing the model's capacity to understand code patterns and relationships among various components.

Additionally, in the evolving landscape of software security, Natural Language Processing (NLP) has emerged as a potent tool with the potential to revolutionize the field. Its application extends to bridging the substantial semantic gap between Natural Language (NL) and programming languages (PL). Nonetheless, a significant semantic disparity exists between Natural Language (NL) and programming languages (PL). To mitigate this divergence and gain a deeper comprehension of the semantic content within NL, scholars introduced CodeBERT \cite{zhangyin2020codebert}, a language model that employs the masked language model and token replacement detection approach to pre-train both NL and PL. CodeBERT has demonstrated remarkable generalization abilities, exhibiting a strong competitive edge in various downstream tasks related to multiple programming languages. The embedding vectors derived from CodeBERT encapsulate a we\-alth of information, thereby enhancing the DL model's ability to yield optimal results post-training.

Therefore, to cope with the above challenges, we propose the XGV-BERT model for better obtaining the contextual representation of programs. Specifically, we leverage a pre-trained model named CodeBERT for code embedding because it is a model pre-trained on multiple programming languages that understand better the source code. Subsequently, the integration of the GNN model with graphs constructed from extracted data helps to enhance the connections between words and slices in source code to disclose the vulnerability in the software program.
In summary, the contributions of our work are illustrated as follows:
\begin{itemize}
    \item We leverage the language model to construct a method of representing vulnerable source code for security defect detection, using CodeBERT embedding model to replace the Word2vec embedding method used in previous studies \cite{li2018vuldeepecker} \cite{li2021sysevr}.
    
    \item We propose a vulnerability detection system called XGV-BERT that utilizes CodeBERT with a Graph Convolutional Network (GCN) model for analyzing C and C++ source code.
    
    \item Our experimental results indicate that XGV-BERT outperforms the state-of-the-art method \cite{li2018vuldeepecker} \cite{li2021sysevr} on the SARD \cite{sard} and NVD \cite{nvd} datasets. 
\end{itemize}

The remaining sections of this article are constructed as follows. Section \ref{related_work} introduced some related works in detecting vulnerabilities in software. In Section \ref{sec_background}, we give the overview of background knowledge used in our work. Next, the proposed framework and methodology are discussed in Section \ref{methodology}. Section \ref{experiments} describes the experimental settings and result analysis of vulnerability detection on various datasets. Finally, we conclude the paper in Section \ref{conclusion}.

\section{Background} \label{sec_background}
\subsection{Abstract Syntax Tree - AST}

\subsubsection{Definition}

In terms of software, an Abstract Syntax Tree (AST) \cite{michael2005ast} embodies a tree-like depiction of the abstract syntax structure inherent in a fragment of text, commonly referred to as source code, authored in a formal language. Each node within this tree serves as a representation of a discernible structure found within the text. More specifics, abstraction in the AST is manifested by not representing every detail present in the actual syntax but focusing on the structure and content-related aspects. For instance, unnecessary single quotes in the syntactic structure are not represented as separate nodes in the tree. Similarly, a syntax structure like the "if" conditional statement can be represented by a single node with three branches. The AST is a vital tool in parsing programming languages. It provides an abstract structural representation of the source code, enabling programs to understand and process code more easily. The abstraction in the AST allows us to concentrate on essential syntax components and overlook irrelevant details. This simplifies the language analysis and processing, while providing a convenient structure for working with and interacting with the source code.

\subsubsection{Design}

The design of the AST is often closely intertwined with the design of the compiler. The core requirements of the design include the following:
\begin{itemize}
    \item Preservation of Variables: Variables must be retained, along with their declaration positions in the source code.
    \item Representation of Execution Order: The order of execution statements must be represented and explicitly determined.
    \item Proper Handling of Binary Operators: The left and right components of binary operators must be stored and accurately determined.
    \item Storage of Identifiers and Assigned Values: The identifiers and their assigned values must be stored within the assignment statements.
\end{itemize}

\subsection{CodeBERT}
CodeBERT \cite{zhangyin2020codebert} is a pre-trained BERT model that combines both natural language (NL) and programming language (PL) encodings to create a comprehensive model suitable for fine-tuning on source code tasks. The model is trained on a large dataset sourced from code repositories and programming documents, leading to improved effectiveness in software program training and source code analysis.

During the pre-training stage, the input data is formed by combining two segments with special separator token: $[CLS]$, $w_1, w_2, ..w_n, [SEP]$, $c_1, c_2, ..., c_m, [EOS]$, with $[CLS]$ is classification token, $[SEP]$ is separator token and $[EOS]$ is "end of the sequence" token. One segment represents natural language text, while the other represents code from a specific programming language. The $[CLS]$ token is a special token placed before the two segments. Following the standard text processing in Transformer, the natural language text is treated as a sequence of words and divided into WordPieces \cite{wu2016google}. A code snippet is regarded as a sequence of tokens. The output of CodeBERT includes (1) contextualized vector representations for each token, encompassing both natural language and code, and (2) the representation of $[CLS]$, serving as a summarized representation.

\subsection{Graph Neural Network - GNN}
\subsubsection{Overview}

A graph is a data structure in computer science comprising two components: nodes and edges \textit{G} = (\textit{V, E}). Each node has edges (\textit{E}) connecting it to other nodes (\textit{V}). A directed graph has arrows on its edges, indicating directional dependencies, while undirected graphs lack such arrows.

Graphs have attracted considerable attention in Machine Le\-arning due to their powerful representational capabilities. Each node is embedded into a vector, establishing its position in the data space. Graph Neural Networks (GNNs) are specialized neural network architectures that operate on graphs.  The primary goal of GNN architecture is to learn an embedding vector containing information about its local neighborhood. This embedding can be used to address various tasks, such as node labeling, node and edge prediction, and more.

In essence, GNNs are a subclass of DL techniques specifically designed for performing inference on graph-structured data. They are applied to graphs and have the ability to perform prediction tasks at the node, edge, and graph levels.

\subsubsection{Classification}

GNNs are divided into three types:
\begin{itemize}
    \item Recurrent Graph Neural Network: In this network, the graph is bidirectional, where data flows in both directions. It applies graph message passing over edges to propagate the output from the initial direction back to the graph nodes, but adjusts edge weights based on the previously applied gradients for that node.
    \item Spectral Convolutional Network: This type shares a similar idea with CNNs. In CNNs, convolution is performed by summing up the values of neighboring data points around a central data point using learnable filters and weights. Spectral-based networks operate on a similar principle, aggregating the attributes of neighboring nodes for a central node. However, spectral-based methods often have higher computational complexity and have gradually been replaced by spatial-based methods.
    \item Spatial Convolutional Network: This approach provides a simpler and more efficient way to handle data. It embeds nodes based on their neighboring nodes. This spatial-based method has become popular due to its simplicity and effectiveness in processing data.
\end{itemize}

\subsection{Graph Convolutional Network - GCN}
GCN (Graph Convolutional Network) \cite{kipf2016gcn} is a powerful neural network architecture designed for machine learning on graphs. In fact, it is so powerful that even a randomly initialized two-layer GCN can produce meaningful feature representations for nodes in the graph.

Specifically, the GCN model takes the graph data \textit{G} = (\textit{V, E}) as input, where:
\begin{itemize}
    \item $N \times F^0$ is the input feature matrix, denoted as \textit{X}, where \textit{N} is the number of nodes, and $F^0$ is the number of input features for each node.
    \item $N \times N$ is the adjacency matrix \textit{A}, representing the structural information about the graph.
\end{itemize}

Thus, a hidden layer in GCN can be written as $H^i$ = $f(H^{i-1}, A)$, where $H^0 = X$, and $f$ is the propagation rule. Each layer $H^i$ corresponds to a feature matrix $N \times F^i$, where each row represents a feature representation of a node. At each layer, these features are aggregated to create the features for the next layer using the propagation rule $f$. This way, the features become increasingly abstract at each consecutive layer. Variants of GCN differ only in the choice of propagation rule $f$.

\section{Related work} \label{related_work}

\subsection{Software Vulnerability}

\subsubsection{Concept}
Software vulnerabilities represent errors, weaknesses, or imperfections within software or operating systems that are susceptible to the influence of attacks or malevolent actions that may inflict harm upon the system or the information it processes. Software vulnerabilities can be exploited by malicious actors to carry out actions such as unauthorized system access, pilfering sensitive information, impeding the normal functioning of the system, or facilitating other forms of attacks.

With the swift development of novel attack techniques, the severity of software vulnerabilities is continuously escalating. All systems inherently harbor latent vulnerabilities; however, the pertinent question remains whether these vulnerabilities are exploited and result in deleterious consequences.

\subsubsection{The current state}
An increasing number of cyberattacks originate from software vulnerabilities, resulting in user data breaches and tarnishing the reputation of companies \cite{lin2020neural}. Despite numerous research efforts proposed to aid in vulnerability detection, vulnerabilities continue to pose a threat to the secure operation of IT infrastructure \cite{daniel2018hacker}. The number of disclosed vulnerabilities in the Common Vulnerabilities and Exposures (CVE) and National Vulnerability Database (NVD) repositories has surged from approximately 4,600 in 2010 to 8,000 in 2014 before skyrocketing to over 17,000 in 2017 \cite{li2018vuldeepecker}, \cite{record2018}. These vulnerabilities may have led to potential threats concerning the secure usage of digital products and devices worldwide \cite{rory2019data}.

\subsubsection{Exploitation Mechanisms of Vulnerabilities}
Upon discovery of a security vulnerability, the attacker can capitalize on it by crafting programs to infiltrate and take control of the targeted device. Once successful in gaining access to the target, attackers may conduct system reconnaissance to familiarize themselves with its workings. Consequently, they can execute diverse actions such as accessing critical files or deploying malicious code. Leveraging such control, attackers can hijack the computer and pilfer data from the victim's device.

Vulnerabilities are sometimes identified either by software developers themselves or through user and researcher alerts. However, in certain cases, hackers or espionage organizations may uncover intrusion techniques but refrain from notifying the developers, leading to so-called "zero-day" vulnerabilities, as developers have not had an opportunity to patch them. As a result, software or hardware remains exposed to threats until patches or fixes are distributed to users.

Software vulnerabilities can lead to grave consequences, granting attackers unauthorized access and control over devices. To obviate such calamities, the detection and remediation of vulnerabilities assume utmost significance. Nevertheless, on certain occasions, vulnerabilities remain latent until they are maliciously leveraged, wreaking considerable havoc upon users. To mitigate risks, regular software and hardware updates to apply patches and fixes are essential.

\subsection{Related research works}

There are myriad research directions employing DL for security vulnerability detection. According to the survey conducted by Peng Zeng and colleagues \cite{peng2020survey}, four primary research avenues are observed:

\begin{itemize}
    \item Utilizing DL Models for Semantic Program Representations: This direction involves the automatic acquisition of semantic program representations using DL models, as proposed by Wang \cite{wang2016semantic}.
    \item Buffer Overflow Vulnerability Prediction from Raw Source Code: Choi's approach \cite{choi2017buffer} entails predicting buffer overflow vulnerabilities directly from raw source code.
    \item Vulnerability Detection for Binary Code: Liu's approach \cite{liu2020binary} targets vulnerability detection within binary code.
    \item Extraction of Vulnerability-Containing Code Patterns from Source Code: Li's methodology \cite{li2018vuldeepecker} revolves around the extraction of vulnerability-containing code patterns from source code to train models.
\end{itemize}

\textit{Direction 1 - Automating Semantic Representation Learning for Vulnerability Prediction}

Wang's research \cite{wang2016semantic} is a pioneering study that employs Deep Belief Networks (DBNs) to delve into the semantic representations of programs. The study's aim is to harness high-level semantic representations learned by neural networks as vulnerability-indicative features. Specifically, it enables the automatic acquisition of features denoting source code vulnerabilities without relying on manual techniques. This approach is not only suited for predicting vulnerabilities within a single project but also for cross-project vulnerability prediction. Abstract Syntax Trees (ASTs) are employed to represent programs as input for DBNs in training data. They proposed a data preprocessing approach, comprising four steps:

\begin{itemize}
    \item Tokenization: The first step involves parsing the source code into tokens.
    \item Token Mapping: The second step maps tokens to integer identifiers.
    \item DBN-based Semantic Feature Generation: The third step employs DBNs to autonomously generate semantic features.
    \item Vulnerability Prediction Model Establishment: The final step utilizes DBNs to establish a vulnerability prediction model.
\end{itemize}

\textit{Direction 2 - End-to-End Buffer Overflow Vulnerability Prediction from Raw Source Code using Neural Networks}

Choi's research \cite{choi2017buffer} stands as the inaugural work providing an end-to-end solution for detecting buffer overflow vulnerabilities. Experimental studies substantiate that neural networks possess the capability to directly learn vulnerability-relevant characteristics from raw source code, obviating the need for code analysis. The proposed neural network is equipped with integrated memory blocks to retain extensive-range code dependencies. Consequently, adapting this network is pivotal in identifying buffer overflow vulnerabilities. Test outcomes demonstrate the method's precision in accurately detecting distinct types of buffer overflow.

However, this approach still harbors limitations, necessitating further enhancements. A primary constraint lies in its inability to identify buffer overflow incidents occurring within external functions, as input data excludes code from external files. Another limitation is the requirement for each line to encompass data assignments for the model to function. Applying this method directly to source code containing conditional statements proves intricate, as attention scores are computed to locate the most relevant code positions.

\textit{Direction 3 - Vulnerability Detection Solution for Binary Code}

Liu's research \cite{liu2020binary} introduces a DL-based vulnerability detection tool for binary code. This tool is developed with the intent of expanding the vulnerability detection domain by mitigating the scarcity of source code. To train the data, binary segments are fed into a Bidirectional Long Short-Term Memory network with Attention mechanism (Att-BiLSTM). The data processing involves three steps:

\begin{itemize}
    \item Initially, binary segments are collected by applying the IDA Pro tool on the original binary code.
    \item In the second step, functions are extracted from binary segments and labeled as "vulnerable" or "non-vulnerable".
    \item In the third step, binary segments are used as binary features before feeding them into the embedding layer of Att-BiLSTM.
\end{itemize}

The granularity of detection lies at the function level. Multiple experiments were conducted on open-source project datasets to evaluate the proposed method. The results of these experiments indicate that the proposed approach outperforms other binary code-based vulnerability detection methods. However, this method still has limitations. Notably, the detection accuracy is relatively low, falling below 80\% in each dataset.

\textit{Direction 4 - Extracting Vulnerable Code Patterns for Model Training}

Li's VulDeePecker \cite{li2018vuldeepecker} is the pioneering study that employs the BiLSTM model for vulnerability detection. This research direction aligns with our team's pursuit as well. This study employs BiLSTM to extract and learn long-range dependencies from code sequences. The training data for the tool is derived from code gadgets representing programs, serving as input for the BiLSTM. The processing of code gadgets involves three stages:

\begin{itemize}
    \item The initial stage entails extracting corresponding program slices of library/API function calls.
    \item The second stage revolves around creating and labeling code gadgets.
    \item The final stage focuses on transforming code gadgets into vectors.
\end{itemize}

Experimental outcomes demonstrate VulDeePecker's capacity to address numerous vulnerabilities, and the integration of human expertise can enhance its effectiveness. However, this method exhibits certain limitations that require further improvement. Firstly, VulDeePecker is restricted to processing programs written in C/C++. Secondly, it can solely address vulnerabilities linked to library/API function calls. Lastly, the evaluation dataset is relatively small-scale, as it only encompasses two types of vulnerabilities.

The reason we opted for the Direction 4 is that we assessed that this research direction offers certain advantages over the other three. In the case of \textit{Direction 1 - Automating Semantic Representation Learning for Vulnerability Prediction}, this research is limited to extracting semantic features from source code alone. Conversely, Direction 4 has seen consecutive studies that extract both semantic and syntactic features, as exemplified by SySeVR \cite{li2021sysevr}, enhancing the reliability of model training data. Regarding \textit{Direction 2 - End-to-End Buffer Overflow Vulnerability Prediction from Raw Source Code using Neural Networks}, it has a drawback in that it exclusively detects Buffer Overflow vulnerabilities. In contrast, Direction 4 employs data extraction methods that can increase the number of vulnerabilities to as many as 126 CWEs, divided into four categories, as we will discuss in Section \ref{experiments}. As for \textit{Direction 3 - Vulnerability Detection Solution for Binary Code}, it holds significant potential as it can detect vulnerabilities in various programming languages by utilizing binary code. However, its accuracy remains lower compared to the other research directions. For these reasons, our team has determined to select Direction 4 as a reference and propose XGV-BERT for further improvement. In our proposed XGV-BERT, the fusion of CodeBERT model and Graph Neural Networks (GNN) represents a compelling strategy within the domain of software vulnerability detection. CodeBERT, an advanced transformer-based model, excels in acquiring meaningful code representations, while GNNs exhibit exceptional prowess in encoding semantic connections present in code graphs. This harmonious combination of CodeBERT and GNNs elevates the precision and efficiency of software vulnerability detection, facilitating the discovery of intricate vulnerabilities that conventional approaches might struggle to uncover.

\section{Methodology} \label{methodology}
\subsection{The overview architecture}

\begin{figure*}[!ht]
\centering
\includegraphics[width=0.85\linewidth]{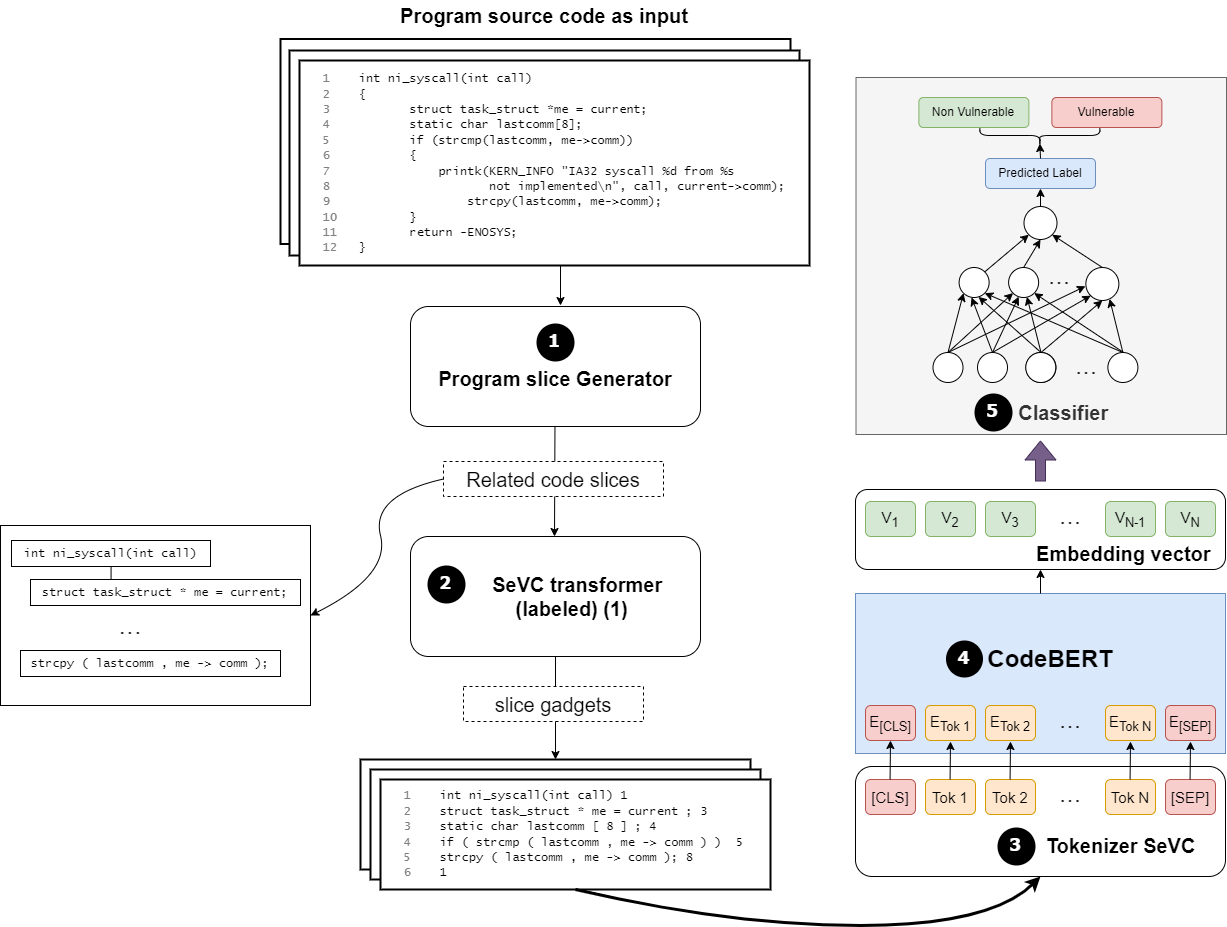}
\caption{The workflow of XGV-BERT framework for vulnerability detection.}
\label{fig:xgvbert}
\end{figure*}

To detect vulnerabilities using DL, we need to represent programs in a way that captures both syntactic and semantic information relevant to the vulnerabilities. There are two main contributions presented in Li's research \cite{li2021sysevr}. To start with, treating each function in the program as a region proposal \cite{ren2017rcnn} similar to image processing. However, this approach is overly simplistic because vulnerability detection tools not only need to determine the existence of vulnerabilities within functions but also need to pinpoint the locations of the vulnerabilities. This means we require detailed representations of the programs to effectively detect vulnerabilities. Secondly, they treat each code line or statement (used interchangeably) as a unit for vulnerability detection. However, their approach has two drawbacks. More specifics, most statements in a program do not contain any vulnerabilities, resulting in a scarcity of vulnerable samples. In addition, many statements have semantic relationships with each other, but they are not considered as a whole.

To synthesize the advantages of the two above proposals, we divide the program into smaller code segments (i.e., a few lines of code), corresponding to region proposals, and represent the syntactic and semantic characteristics of the vulnerabilities.

After studying Li's method \cite{li2021sysevr}, our research group proposes the XGV-BERT method for efficiently detecting software vulnerability. We extract feature patterns from the source code using the CodeBERT model to embed the vectors, and then we feed them into various DL models, including RNN, LSTM, Bi-LSTM, GRU, Bi-GRU, and the proposed GCN model for comparison and evaluation. Figure \ref{fig:xgvbert} illustrates the specific architecture of the steps involved in our proposed software vulnerability detection method.

The architecture we propose uses the original source code as input, followed by the extraction of program slices based on syntax features. From the program slices, we further extract lines of code that have semantic relationships with each program slice, creating Semantics-based Vulnerability Candidates (SeVCs) \cite{li2021sysevr} or code gadgets \cite{li2018vuldeepecker}, depending on the dataset, and label them accordingly. Then, we tokenize the SeVCs or code gadgets and feed them into the CodeBERT model for vector embedding. Finally, we construct a DL model for training and predicting vulnerable source code using the embedded vectors.

\subsection{Embedding Vectors}
\label{token-sevc}
In this section, we delve into the process of tokenizing and embedding the SeVCs extracted from the source code for training in the DL model. The following steps outline our approach:

\begin{itemize}
    \item Symbolic Representation of SeVCs: To ensure the independence of SeVCs from user-defined variables and function names while capturing the program's semantic information, each SeVC undergoes transformation into a symbolic representation.
    \begin{itemize}
        \item Removal of non-ASCII characters and comments from the code.
        \item Mapping of user-defined variable names to symbolic names (e.g., "V1", "V2") in a one-to-one correspondence.
        \item Mapping of user-defined function names to symbolic names (e.g., "F1", "F2") in a one-to-one correspondence.
    \end{itemize}
    It is important to note that different SeVCs may share the same symbolic representation, enhancing the generalizability of the approach.
    \item Tokenize the symbolic representations: To achieve this, Professor Li's team \cite{li2021sysevr} proposed to split the symbol representation of SeVC (e.g., "V1=V2-8;") into a sequence of symbols through lexical analysis (e.g., "V1", "=", "V2", "-", "8", and ";"). Each symbol in the resulting sequence is considered a token. This process is performed for each code line in the SeVC, resulting in a list of tokens for each SeVC.
    \item After obtaining the list of tokens, we use the CodeBERT model to embed the data. The CodeBERT model used in this study has been pretrained on source code data and is retrained with the input of the tokenized vectors. The model architecture consists of multiple Transformer layers, which extract features from the dataset and enhance contextual information for the vectors. The output of the model will be the embedded vectors, which we use as inputs for the DL models to classify the dataset.
\end{itemize}

\subsection{Training DL Models}
In the final part, we utilize DL models for training and classifying the dataset. Specifically, we employ a total of 6 models: RNN, LSTM, Bi-LSTM, GRU, Bi-GRU, and GNN. The input to these models is the embedded vectors. Among these models, the most significant one we propose is GNN, named XGV-BERT.

For models such as RNN and its variants, including LSTM, GRU, Bi-LSTM, and Bi-GRU, we implement these models with inputs being the embedded vectors. The architecture of these models consists of two hidden layers, accompanied by fully connected layers to converge the output for text classification purposes.

\begin{figure*}[!ht]
\centering
\includegraphics[width=0.85\linewidth]{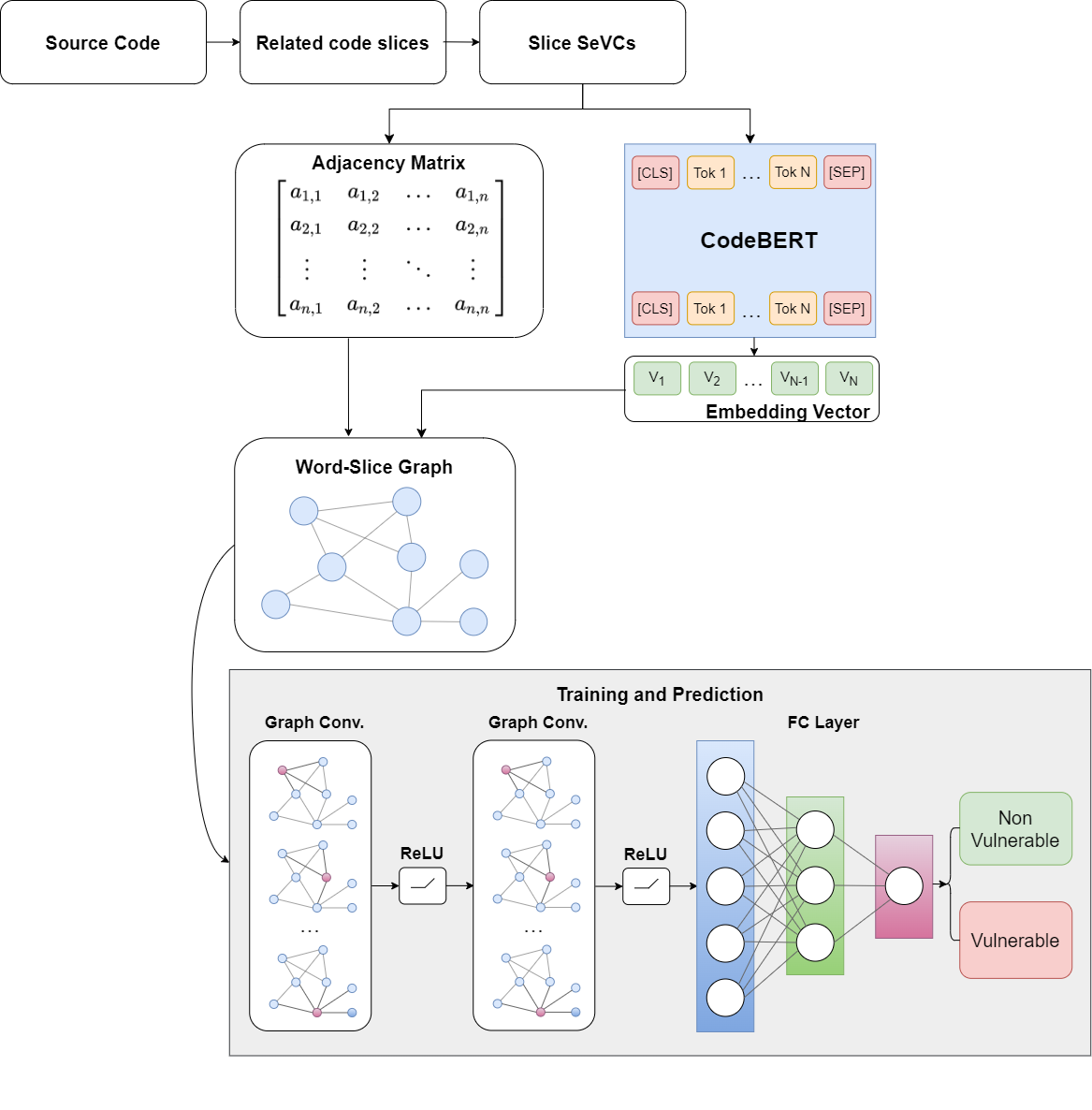}
\caption{The proposed GCN model in XGV-BERT framework.}
\label{fig:diagram}
\end{figure*}

For the GNN model, we employ the GCN architecture for training. Our GCN model is designed to take input data in the form of graphs. The embedded vectors obtained in section \ref{token-sevc} need to be processed into graph-structured data before being fed into our training model. To accomplish this, we create adjacency matrices for the embedded vectors, and these vectors are transformed into graph nodes. Figure \ref{fig:diagram} illustrates the architecture using the proposed GCN model for training and predicting results on the dataset.

Specifically, we construct a non-homogeneous graph consisting of both word nodes and slice nodes based on the idea proposed by Yao \cite{yao2019textgcn}. In this adjacency matrix, each word or slice is represented as a one-hot vector and used as input for the GCN model. We create edges between nodes based on the occurrence of a word in a slice (slice-word edge) and the co-occurrence of words across the entire dataset (word-word edge). The weight of an edge between two nodes \textit{i} and \textit{j} is defined as follows:

\[ A_{i,j} =
    \begin{cases}
    \text{PPMI}(i,j), \quad i,j \text{ are words and } i \neq j\\
    \text{TF-IDF}(i,j), \quad i \text{ is slice, } j \text{ is word}\\
    1, \quad i = j\\
    0, \quad \text{otherwise}
    \end{cases}
\]

The Term Frequency-Inverse Document Frequency (TF-IDF) value of a word in a slice determines the weight of the edge between a slice node and a word node. This value is used to assess the importance of a word in a slice, where a higher value indicates a higher importance of the word for the slice. Specifically:

\begin{itemize}
    \item TF (Term Frequency) is the number of times the word appears in the slice.
    \item IDF (Inverse Document Frequency) is calculated as follows:
    \begin{equation}
        \text{IDF} = \log\frac{N}{n_{t}}
    \end{equation}
    Where $N$ is the total number of slices in the dataset, and ${n_{t}}$ is the number of slices that contain the word. IDF helps evaluate the importance of a word for a slice, assigning lower IDF scores to frequently occurring words.
    \item The TF-IDF weight is computed by multiplying the Term Frequency (TF) and the Inverse Document Frequency (IDF).
\end{itemize}

The Positive Pointwise Mutual Information (PPMI) is used to determine the weight of word pairs, where a higher PPMI value indicates a stronger relationship and co-occurrence frequency between two words. The PPMI value of $i$ and $j$ is calculated using the formula:
\begin{equation}
    \text{PPMI}(i,j) = \max (\log\frac{p(i,j)}{p(i)p(j)},0)
\end{equation}
Where:

\begin{itemize}
    \item $p(i)$ is the probability of word $i$ appearing in a slice.
    \item $p(j)$ is the probability of word $j$ appearing in a slice.
    \item $p(i,j)$ is the joint probability of both $i$ and $j$ appearing in a slice.
\end{itemize}

For the nodes in the adjacency matrix, we utilize the output embeddings of the CodeBERT model and use them as input representations for the slice nodes. Once the data, including the adjacency matrix and the nodes, is constructed, we utilize both of them as inputs for the GCN model. Our GCN model consists of two hidden layers, along with fully connected layers, to converge the output for text classification purposes.

\section{Experiments and Analysis} \label{experiments}

In this section, we conduct experiments to compare the XGV-BERT's detection accuracy to the state-of-the-art solutions, VulDeePecker \cite{li2018vuldeepecker} and SySeVR \cite{li2021sysevr}. Before discussing the effectiveness of XGV-BERT, we first go over the implementation specifics and datasets used in the trials.

\subsection{Dataset and Preprocessing}
\subsubsection{Benchmark Dataset}

For the evaluation of our approach, we use two datasets from two research papers: SySeVR \cite{li2021sysevr} and VulDeePecker \cite{li2018vuldeepecker}. Both datasets were collected from two sources: the National Vulnerability Database (NVD) \cite{nvd} and the Software Assurance Reference Dataset (SARD) \cite{sard}.

The NVD dataset provides vulnerable code snippets from various software products, including both vulnerable code and their corresponding patches. On the other hand, the SARD dataset offers a collection of test, synthetic, and academic programs, labeled as "good" (having no vulnerabilities), "bad" (containing vulnerabilities), and "mixed" (having vulnerabilities whose
patched versions are also available).

Regarding the VulDeePecker dataset, they offer program pieces that concentrate on two categories of CWE related to Library/API Function Call vulnerabilities, including resource management error vulnerabilities (CWE-399) and buffer error vulnerabilities (CWE-119). We produced 29,313 non-vulnerable code gadgets and 10,440 vulnerable code gadgets for CWE-119. We produced 10,440 vulnerable and 14,600 non-vulnerable code gadgets for CWE-399. The number of code gadgets that we extracted from the VulDeePecker dataset is shown in Table \ref{tab:result-label-vdpecker}.

The specific steps for code gadget extraction are as follows:

\begin{itemize}
    \item Extracting library/API function calls and their corresponding program slices:
    \begin{itemize}
        \item Categorize library/API function calls into two types: forward library/API function calls and backward library/API function calls. The forward type receives inputs directly from external sources, like a command line, program, or file. The backward type does not receive direct inputs from external sources or the program environment.
        \item Generate program slices corresponding to the arguments of the library/API function calls extracted in the previous step. Program slices are further classified into forward slices and backward slices. Forward slices contain statements affected by specific arguments, while backward slices comprise statements influencing specific arguments.
    \end{itemize}
    \item Extracting code gadgets and assigning labels:
    \begin{itemize}
        \item Extract code gadgets:
        \begin{itemize}
            \item Construct a part of the code gadget by combining statements containing arguments from the library/API function calls. These statements belong to the same user-defined function and are ordered based on the corresponding program slice. Any duplicate statements are eliminated.
            \item Complete the code gadget by incorporating statements from other functions containing the arguments from the library/API function calls, following the order in the corresponding program slice.
        \end{itemize}
        \item Label the code gadgets: Those without vulnerabilities receive the label "0," while those containing vulnerabilities are labeled "1."
    \end{itemize}
\end{itemize}

\begin{table}[h]
    \centering
    \caption{Number of code gadgets extracted from the VulDeePecker dataset}
    \label{tab:result-label-vdpecker}
    \scalebox{1}{
    \begin{tabular}{|c|c|c|c|}
    \hline
    Dataset & Total & \multicolumn{1}{m{1.5cm}|}{Vulnerable code gadgets} & \multicolumn{1}{m{1.8cm}|}{Non-vulnerable code gadgets} \\
    \hline
    CWE-119 & 39,753 & 10,440 & 29,313 \\
    \hline
    CWE-399 & 21,885 & 7,285 & 14,600 \\
    \hline
    All Dataset & 61,637 & 17,725 & 43,913 \\
    \hline
    \end{tabular}}
\end{table}

For the SySeVR dataset, it provides C/C++ programs containing 126 CWE related to four types of vulnerabilities: Library/API Function Call (FC-kind), Array Usage (AU-kind), Pointer Usage (PU-kind), and Arithmetic Expression (AE-kind). In total, we have extracted 547,347 SeVCs from the dataset, comprising 52,227 SeVCs containing vulnerabilities and 495,120 SeVCs without vulnerabilities. The distribution of SeVCs based on vulnerability types and their corresponding CWE identifiers is presented in Table \ref{tab:result-label-sevc}.

\begin{itemize}
    \item Step 1. Extract Syntax-based Vulnerability Candidates (SyVCs).
    \begin{itemize}
        \item Represent each function as an Abstract Syntax Tree (AST). The root of the AST corresponds to the function, the leaves represent the tokens of the function, and the intermediate nodes correspond to the statements of the function.
        \item Compare the code elements (comprising one or more consecutive tokens, including identifiers, operators, constants, and keywords) in the AST with a set of syntactic vulnerability patterns. If a code element matches any element in this set, it becomes a SyVC.
    \end{itemize}
    \item Step 2. Transform SyVCs into Semantic-based Vulnerability Candidates (SeVCs).
    \begin{itemize}
        \item Create CFGs for each function in the program. From CFGs, generate PDGs for each function.
        \item Based on the PDG, create program slices ps$_i$ for each SyVC.
        \begin{itemize}
            \item Interprocedural forward slice fs'$_i$ is formed by merging the forward slice fs$_i$ of function $f_i$ with the forward slices of functions called by $f_i$.
            \item Interprocedural backward slice bs'$_i$ is formed by merging the backward slice bs$_i$ of function $f_i$ with the backward slices of functions called by $f_i$ and the functions that call $f_i$.
            \item Finally, program slice ps$_i$ is created by merging fs'$_i$ and bs'$_i$.
        \end{itemize}
        \item Transform the program slice into SeVCs with the following steps:
        \begin{itemize}
            \item Convert the statements belonging to function $f_i$ and appearing in program slice ps$_i$ as a node into SeVCs while preserving the original order of statements in function $f_i$.
            \item Convert the statements belonging to other functions, which are related to function $f_i$ through function calls, into SeVCs.
        \end{itemize}
    \end{itemize}
    \item Step 3. Label the SeVCs: To differentiate between vulnerable and safe code patterns, we label the SeVCs, and their corresponding vectors, accordingly. A SeVC containing a known vulnerability is labeled as "1," while it is labeled as "0" if the SeVC is safe and does not contain any vulnerabilities.
\end{itemize}

\begin{table}[h]
    \centering
    \caption{Number of SeVCs extracted from the SySeVR dataset}
    \label{tab:result-label-sevc}
    \scalebox{1}{
    \begin{tabular}{|c|c|c|c|}
    \hline
    Dataset & Total & \multicolumn{1}{m{1.2cm}|}{Vulnerable SeVCs} & \multicolumn{1}{m{1.8cm}|}{Non-vulnerable SeVCs} \\
    \hline
    FC-kind & 141,023 & 17,005 & 124,018 \\
    \hline
    AU-kind  &  55,772  &  7,996 & 47,776 \\
    \hline
    PU-kind   &  340.324 & 25.377 & 314.946 \\
    \hline
    AE-kind & 10.234 & 1.848 & 8.386 \\
    \hline
    All Dataset & 61,637 & 17,725 & 43,913 \\
    \hline
    \end{tabular}}
\end{table}

By leveraging these datasets, we were able to comprehensively evaluate the effectiveness and performance of our proposed method in detecting software vulnerabilities.

\subsection{Performance Metrics} \label{sec:performance_metric}
\subsubsection{Detection metrics}

To accordingly evaluate the model prediction, we discussed and defined ground truth values as follows: true positive (TP) represents the number of vulnerable
samples that are detected as vulnerable; true negative (TN) represents the number of samples that are not vulnerable and are detected as not vulnerable; False positive (FP) represents the number of samples are not vulnerable but are detected
as vulnerable; False negative (FN) represents the number of vulnerable samples that are detected as not vulnerable.

Therefore, we use four metrics as follows for our experiments:

\begin{itemize}
    \item \emph{Accuracy} is the ratio of correct and total predictions.
    \begin{equation} \label{eq:accuracy}
        Accuracy = \frac{TP + TN}{TP + TN + FP + FN}
    \end{equation}
    \item \emph{Precision} is the ratio of truly vulnerable samples among the detected vulnerable samples.
    \begin{equation} \label{eq:precision}
        Precision = \frac{TP}{TP + FP}
    \end{equation}
    \item \emph{Recall} is the proportion of truly vulnerable samples among the samples that were predicted as not containing vulnerabilities.
    \begin{equation} \label{eq:recall}
        Recall = \frac{TP}{TP + FN}
    \end{equation}
    \item \emph{F1-score} measures the overall effectiveness by calculating the harmonic mean of Precision and Recall.
    \begin{equation} \label{eq:f1-score}
        F1-score = 2 \cdot \frac{Recall \cdot Precision}{Recall + Precision}
    \end{equation}
\end{itemize}

\subsection{Experimental Settings} \label{sec:experimental_setting}

We conducted our experiments on a virtual machine environment running Ubuntu 20.04  with a 8 core CPUs, 81.5GB of RAM, 40GB of GPU RAM, and a storage capacity of 100GB. TABLE \ref{tab:bilstm-model-architecture} and TABLE \ref{tab:xgv-bert-model-architecture} show the architecture for the LSTM model and XGV-BERT, respectively. To perform our experiments, we trained the datasets using the following configuration on both models: $Adam\ Optimizer$ with $learning\_rate=0.001$, $epoch=50$ and $ batch\_size=32$ for RNN, LSTM, BiLSTM, GRU and BiGRU models and the same configuration for XGV-BERT model, but with $epoch=4$. In the settings for both VulDeePecker and SySeVR datasets, we choose 80\% samples for the training set and the remaining 20\% for the test set.

\begin{table}[ht]
    \centering
    \caption{The architecture of CNN\&GRU model}
    \label{tab:bilstm-model-architecture}
    \scalebox{1}{
    \begin{tabular}{|c|c|c|c|}
    \hline
    Layer (ID) & Activation & Output shape & Connected to  \\
    \hline
    Input (1) & - & (768, 1) & [] \\
    \hline
    LSTM (2) & ReLU & (768, 300) & [(1)] \\
    \hline
    Dropout (3) & - & (768, 300) & [(2)] \\
    \hline
    LSTM (4) & ReLU & (300) & [(3)] \\
    \hline
    Dropout (5) & - & (300) & [(4)] \\
    \hline
    Dense (6) & Sigmoid & (1) & [(5)] \\
    \hline
    \end{tabular}}
\end{table}

\begin{table}[h]
    \centering
    \caption{The architecture of XGV-BERT model}
    \label{tab:xgv-bert-model-architecture}
    \scalebox{1}{
    \begin{tabular}{|c|c|c|c|}
    \hline
    Layer (ID) & Activation & Output shape & \makecell{Connected \\ to} \\
    \hline
    Input (1) & - & \makecell{[Node: (1, 768), \\ Edge: (1, 768)]} & [] \\
    \hline
    \makecell{GraphConv \\ (2)} & ReLU & Node: (1, 200) & [(1)] \\
    \hline
    \makecell{GraphConv \\ (3)} & ReLU & Node: (1, 200) & [(2)] \\
    \hline
    Dropout (4) & - & Node: (1, 200) & [(3)] \\
    \hline
    Dropout (5) & - & Node: (1, 200) & [(4)] \\
    \hline
    Dense (6) & - & Edge: (2) & [(5)] \\
    \hline
    \end{tabular}}
\end{table}

\subsection{Experimental Results}
\subsubsection{Effectiveness of CodeBERT embedding method}
To demonstrate the effectiveness of our embedding method using the CodeBERT model (see Section \ref{token-sevc}), we conducted experiments and evaluated the performance of the  Word2vec and CodeBERT embedding models. We compared these results to the experimental findings of two related research studies, SySeVR \cite{li2021sysevr} and VulDeePecker \cite{li2018vuldeepecker}. Tables ~\ref{tab:result-vdpk-a-classification} and ~\ref{tab:result-sevc-a-classification} summarize the evaluation's findings.

\begin{table}[htbp]
\centering
\caption{Evaluation results of vector embedding methods on VulDeePecker dataset}
\begin{tabular}{|l|c|c|c|c|}
\hline
Training method & Acc & Precision & Recall & F1 \\
\hline
VulDeePecker (BiLSTM) & \textbf{90.8} & 79.1 & 77.5 & 78.3 \\
Our Word2Vec+LSTM & 82.6 & 85.4 & 82.3 & 83.8 \\
Our Word2Vec+BiLSTM & 83.4 & 86.9 & 81.9 & 84.3 \\
CodeBERT+LSTM & 83.5 & 80.9 & \textbf{87.7} & 84.2 \\
CodeBERT+BiLSTM & 86.0 & \textbf{89.5} & 85.3 & \textbf{87.3} \\
\hline
\end{tabular}

\label{tab:result-vdpk-a-classification}
\end{table}

Table ~\ref{tab:result-vdpk-a-classification} demonstrates that our embedding method using the CodeBERT model has achieved significantly better evaluation results overall compared to VulDeePecker \cite{li2018vuldeepecker}. The F1 score for our approach is 87.3\%, representing a substantial improvement over VulDeePecker's F1 score of 79.3\%. Similarly, the Precision and Recall metrics have also shown improvements compared to VulDeePecker. Although the detection accuracy when we used CodeBERT (86.0\%) was not superior to VulDeepecker (90.8\%), it showed improvement when compared with our experimental evaluation using Word2vec (83.4\%). In these cases, the F1 metric is more suitable for evaluating the performance of vulnerability detection because of the minority of vulnerable code compared to clean code.

\begin{table}[htbp]
\centering
\caption{Evaluation results of vector embedding methods on SySeVR dataset}
\begin{tabular}{|l|c|c|c|c|}
\hline
Training method & Acc & Precision & Recall & F1 \\
\hline
SySeVR (LSTM) & 95.2 & 85.2 & 78.3 & 81.6 \\
SySeVR (BiLSTM) & \textbf{96.0} & 86.2 & 82.5 & 84.3 \\
Our Word2Vec+LSTM & 91.8 & 89.8 & 81.1 & 85.2 \\
Our Word2Vec+BiLSTM & 92.6 & \textbf{92.1} & 81.5 & 86.4 \\
CodeBERT+LSTM & 93.3 & 88.3 & \textbf{99.8} & 93.7\\
CodeBERT+BiLSTM & 93.8 & 89.1 & 99.7 & \textbf{94.1} \\
\hline
\end{tabular}
\label{tab:result-sevc-a-classification}
\end{table}

Meanwhile, Table ~\ref{tab:result-sevc-a-classification} shows the experimental results on the SySeVR dataset \cite{li2021sysevr}. Again, our embedding method using CodeBERT outperforms the embedding method using SySeVR's Word2vec. By combining BiLSTM with CodeBERT embeddings, we achieved a significant improvement in the F1-score, increasing it from 84.3\% to 94.1\% when compared to SySeVR.

\subsubsection{Effectiveness of DL models}
\begin{table*}[h]
\centering
\caption{Test performance of various models using CodeBERT embedding method on VulDeePecker dataset}
\label{tab:result-vdpk-classification}
\begin{tabular}{ |c|c|c|c|c|c|c|c| } 
\hline
Dataset & Metrics & RNN & LSTM & BiLSTM & GRU & BGRU & XGV-BERT \\ \hline
& Accuracy & 69.3 & 77.4 & 79.4 & 78.5 & 79.1 & \textbf{98.4}\\

CWE & Precision & 63.6 & 70.0 & 73.3 & 71.5 & 75.0 & \textbf{97.9}\\

119 & Recall & 90.3 & 96.0 & 92.6 & 94.6 & 87.5 & \textbf{98.1}\\

& F1-score & 74.6 & 81.0 & 81.8 & 81.5 & 80.8 & \textbf{97.7}\\
\hline

 & Accuracy & 80.6 & 89.0 & 91.0 & 89.3 & 91.0 & \textbf{98.3}\\

CWE & Precision & 72.3 & 84.5 & 92.4 & 86.5 & 91.7 & \textbf{97.9}\\

399 & Recall & 72.3 & 84.5 & 92.4 & 86.5 & 92.4 & \textbf{98.1} \\

& F1-score & 82.4 & 88.5 & 91.1 & 89.0 & 91.2 & \textbf{98.0} \\
\hline

& Accuracy & 76.8 & 83.5 & 86.0 & 82.4 & 86.0 & \textbf{97.8} \\

All & Precision & 74.1 & 80.9 & 89.5 & 79.6 & 81.6 & \textbf{97.3} \\

Dataset & Recall & 82.4 & 87.7 & 85.3 & 87.2 & 94.4 & \textbf{97.7} \\

& F1-score & 78.1 & 84.2 & 87.3 & 83.2 & 87.6 & \textbf{97.5} \\
\hline
\end{tabular}
\end{table*}

\begin{table*}[h]
\centering
\caption{Test performance of various models using CodeBERT embedding method on SySeVR dataset }
\label{tab:result-sysevr-classification}
\begin{tabular}{ |c|c|c|c|c|c|c|c| } 
\hline
Dataset & Metrics & RNN & LSTM & BiLSTM & GRU & BGRU & XGV-BERT \\ \hline
& Accuracy & 88.4 & 91.5 & 93.0 & 92.1 & 92.8 & \textbf{97.2}\\

Library/API & Precision & 81.8 & 85.6 & 87.9 & 86.5 & 87.5 & \textbf{94.4}\\

Function Call & Recall & \textbf{99.8} & 99.7 & 99.7 & 99.7 & 99.7 & 92.5\\

& F1-score & 89.5 & 92.1 & \textbf{93.4} & 92.7 & 93.2 & \textbf{93.4}\\
\hline

& Accuracy & 87.3 & 91.3 & 91.9 & 91.1 & 92.2 & \textbf{98.6}\\

Array & Precision & 79.9 & 85.4 & 86.4 & 94.9 & 86.6 & \textbf{96.7}\\

Usage & Recall & \textbf{99.8} & 99.7 & 99.6 & 99.9 & 99.7 & 97.9\\

& F1-score & 88.7 & 92.0 & 92.5 & 91.8 & 92.7 & \textbf{97.3}\\
\hline

& Accuracy & 89.4 & 94.2 & 94.6 & 94.0 & 94.6 & \textbf{99.7}\\

Pointer & Precision & 82.7 & 89.8 & 90.7 & 89.6 & 90.6 & \textbf{98.5}\\

Usage & Recall & 99.6 & \textbf{99.7} & 99.4 & 99.6 & 99.5 & 97.5\\

& F1-score & 90.4 & 94.5 & 94.9 & 94.3 & 94.9 & \textbf{98.0}\\
\hline

& Accuracy & 78.9 & 84.5 & 88.7 & 85.4 & 87.6 & \textbf{95.5}\\

Arithmetic & Precision & 72.1 & 76.5 & 81.6 & 77.5 & 80.6 & \textbf{90.1}\\

Expression & Recall & 94.3 & 99.5 & \textbf{99.7} & \textbf{99.7} & 98.9 & 96.3\\

& F1-score & 81.7 & 86.5 & 89.8 & 87.2 & 88.8 & \textbf{92.8}\\
\hline

& Accuracy & 88.4 & 93.3 & 93.8 & 93.0 & 93.9 & \textbf{97.8}\\

All & Precision & 81.4 & 88.3 & 89.1 & 87.8 & 89.4 & \textbf{94.8}\\

Dataset & Recall & 99.7 & \textbf{99.8} & 99.7 & 99.7 & 99.7 & 96.2\\

& F1-score & 89.6 & 93.7 & 94.1 & 93.4 & 94.3 & \textbf{95.5}\\
\hline
\end{tabular}
\end{table*}

The TABLE~\ref{tab:result-vdpk-classification} and TABLE~\ref{tab:result-sysevr-classification} presents the detection performance metrics of six models on the VulDeePecker \cite{li2018vuldeepecker} and SySeVR \cite{li2021sysevr} datasets, where all models use CodeBERT's embedding vectors as input. Overall, all models performed exceptionally well, exhibiting high scores across all evaluation metrics. Notably, for the VulDeePecker dataset, our proposed XGV-BERT method achieved the highest rating on all four indices, outperforming the remaining five models. Similarly, for the SySeVR dataset, XGV-BERT achieved the highest scores on three out of four metrics, including accuracy, precision and F1-score. Based on this evaluation result, we can see that XGV-BERT gives the best classifier performance in the compared DL models. These results indicate that XGV-BERT, which leverages the CodeBERT and GNN, can represent the contextual data to identify the vulnerable code in the software with high performance. In summary, the integration of the CodeBERT model with GNN has proven as a promising approach in the realm of software vulnerability detection. CodeBERT, a pre-trained Transformer-based model, excels at learning representations of source code, while GNNs possess a remarkable capability to capture semantic relationships within code graphs. This synergy between CodeBERT and GNNs enhances the accuracy and efficacy of software vulnerability detection, enabling the identification of complex vulnerabilities that may remain elusive through conventional methods.

\section{Conclusion} \label{conclusion}

In concluding, this study introduces a novel method employing contextual embedding and deep learning techniques for the classification of software programs with vulnerabilities. The newly devised framework, termed XGV-BERT, leverages the sophisticated capabilities of contextualized embeddings through CodeBERT to delineate the intricate interconnections among code attributes essential for identifying security flaws. Within the realm of source code analysis, such embeddings are pivotal in discerning the nuanced relationships between tokens or words present in code fragments. Such embeddings empower the model to depict this variable uniquely, contingent on its positional context. Furthermore, XGV-BERT integrates CodeBERT with the advanced Graph Convolutional Network (GCN) deep learning paradigm. A salient feature of GCNs is their adeptness at assimilating contextual intelligence from elaborate graph formations. These networks intrinsically evolve context-sensitive attributes, obviating the necessity for labor-intensive feature crafting. Significantly, GCNs excel in identifying multi-layered contextual associations by analyzing not only the immediate context of a given entity but also the surrounding environment of its neighboring entities and their interconnections. This intrinsic property renders GCNs exceptionally equipped for apprehending multifaceted dependencies within graph-centric data, thereby bolstering their utility across diverse applications. Such an amalgamation augments the depth of learning and information extraction from the multifarious segments inherent in source codes.

Our framework can help cybersecurity experts detect errors and vulnerabilities in software programs automatically with high accuracy. The experimental results on the two benchmark datasets, including VulDeePecker and SySeVR, demonstrate the effectiveness of the proposed framework in improving the performance and detection accuracy of DL-based vulnerability detection systems. In the future, we aim to enhance the source code extraction framework. Our primary objective is to refine vulnerability detection granularity. At present, our system operates at the slice-level, focusing on multiple semantically interrelated lines of code. Additionally, we aspire to expand our vulnerability detection capabilities across diverse programming languages, as the current framework is limited to extracting information solely from C/C++ source code.


%

\section*{Acknowledgment}


This research was supported by The VNUHCM-University of Information Technology's Scientific Research Support Fund.


 \bibliographystyle{elsarticle-num} 
 \bibliography{cas-refs}

\begin{thebibliography}{10}
\expandafter\ifx\csname url\endcsname\relax
  \def\url#1{\texttt{#1}}\fi
\expandafter\ifx\csname urlprefix\endcsname\relax\def\urlprefix{URL }\fi
\expandafter\ifx\csname href\endcsname\relax
  \def\href#1#2{#2} \def\path#1{#1}\fi

\bibitem{zhu2023application_SoftSec_Review}
Y.~Zhu, G.~Lin, L.~Song, J.~Zhang, The application of neural network for software vulnerability detection: a review, Neural Computing and Applications 35~(2) (2023) 1279--1301.

\bibitem{cve}
\href{https://cve.mitre.org/}{Common vulnerabilities exposures (cve)}.
\newline\urlprefix\url{https://cve.mitre.org/}

\bibitem{yonghee2010}
Y.~Shin, A.~Meneely, L.~Williams, J.~A.~Osborne, Evaluating complexity, code churn, and developer activity metrics as indicators of software vulnerabilities, IEEE Transactions on Software Engineering, vol. 37, no. 6 (2010) 772--787.

\bibitem{zeng2020software_DL_sur}
P.~Zeng, G.~Lin, L.~Pan, Y.~Tai, J.~Zhang, Software vulnerability analysis and discovery using deep learning techniques: A survey, IEEE Access 8 (2020) 197158--197172.

\bibitem{hanif2021riseML_soft}
H.~Hanif, M.~H. N.~M. Nasir, M.~F. Ab~Razak, A.~Firdaus, N.~B. Anuar, The rise of software vulnerability: Taxonomy of software vulnerabilities detection and machine learning approaches, Journal of Network and Computer Applications 179 (2021) 103009.

\bibitem{le2022survey}
T.~H. Le, H.~Chen, M.~A. Babar, A survey on data-driven software vulnerability assessment and prioritization, ACM Computing Surveys 55~(5) (2022) 1--39.

\bibitem{marjanov2022machine}
T.~Marjanov, I.~Pashchenko, F.~Massacci, Machine learning for source code vulnerability detection: What works and what isn’t there yet, IEEE Security \& Privacy 20~(5) (2022) 60--76.

\bibitem{lin2020software}
G.~Lin, S.~Wen, Q.-L. Han, J.~Zhang, Y.~Xiang, Software vulnerability detection using deep neural networks: a survey, Proceedings of the IEEE 108~(10) (2020) 1825--1848.

\bibitem{khan2021systematic}
R.~A. Khan, S.~U. Khan, H.~U. Khan, M.~Ilyas, Systematic mapping study on security approaches in secure software engineering, Ieee Access 9 (2021) 19139--19160.

\bibitem{zou2020muy}
D.~Zou, S.~Wang, S.~Xu, Z.~Li, H.~Jin, "$\mu$vuldeepecker: A deep learning-based system for multiclass vulnerability detection", IEEE Transactions on Dependable and Secure Computing 18~(5) (2021) 2224--2236.
\newblock \href {https://doi.org/10.1109/TDSC.2019.2942930} {\path{doi:10.1109/TDSC.2019.2942930}}.

\bibitem{alec2018pretrain}
A.~Radford, K.~Narasimhan, T.~Salimans, I.~Sutskever, Improving language understanding by generative pre-training (2018).

\bibitem{jacob2018bert}
J.~Devlin, M.-W. Chang, K.~Lee, K.~Toutanova, Bert: Pre-training of deep bidirectional transformers for language understanding, arXiv preprint arXiv:1810.04805 (2018).

\bibitem{zhangyin2020codebert}
Z.~Feng, D.~Guo, D.~Tang, N.~Duan, X.~Feng, M.~Gong, L.~Shou, B.~Qin, T.~Liu, D.~Jiang, M.~Zhou, \href{https://aclanthology.org/2020.findings-emnlp.139}{{C}ode{BERT}: A pre-trained model for programming and natural languages}, in: Findings of the Association for Computational Linguistics: EMNLP 2020, Association for Computational Linguistics, Online, 2020, pp. 1536--1547.
\newblock \href {https://doi.org/10.18653/v1/2020.findings-emnlp.139} {\path{doi:10.18653/v1/2020.findings-emnlp.139}}.
\newline\urlprefix\url{https://aclanthology.org/2020.findings-emnlp.139}

\bibitem{peng2020survey}
P.~Zeng, G.~Lin, L.~Pan, Y.~Tai, J.~Zhang, Software vulnerability analysis and discovery using deep learning techniques: A survey, IEEE Access (2020).

\bibitem{wang2016semantic}
S.~Wang, T.~Liu, L.~Tan, Automatically learning semantic features for defect prediction, Proc. 38th Int. Conf. Softw. Eng. (ICSE) (2016) 297--308.

\bibitem{choi2017buffer}
M.-j. Choi, S.~Jeong, H.~Oh, J.~Choo, End-to-end prediction of buffer overruns from raw source code via neural memory networks, arXiv preprint arXiv:1703.02458 (2017).

\bibitem{li2018vuldeepecker}
Z.~Li, D.~Zou, S.~Xu, X.~Ou, H.~Jin, S.~Wang, Z.~Deng, Y.~Zhong, Vuldeepecker: A deep learning-based system for vulnerability detection, Proc. Netw. Distrib. Syst. Secur. Symp. (2018).

\bibitem{liu2020binary}
S.~Liu, M.~Dibaei, Y.~Tai, C.~Chen, J.~Zhang, Y.~Xiang, Cyber vulnerability intelligence for internet of things binary, IEEE Trans. Ind. Informat. (2020) 2154--2163.

\bibitem{croft2022data_softsec_data_sur}
R.~Croft, Y.~Xie, M.~A. Babar, Data preparation for software vulnerability prediction: A systematic literature review, IEEE Transactions on Software Engineering 49~(3) (2022) 1044--1063.

\bibitem{li2021sysevr}
Z.~Li, D.~Zou, S.~Xu, H.~Jin, Y.~Zhu, Z.~Chen, Sysevr: A framework for using deep learning to detect software vulnerabilities, arXiv preprint arXiv:1807.06756 (2021).

\bibitem{kim2022vuldebert}
S.~Kim, J.~Choi, M.~E. Ahmed, S.~Nepal, H.~Kim, Vuldebert: A vulnerability detection system using bert, 2022 IEEE International Symposium on Software Reliability Engineering Workshops (ISSREW) (2022).

\bibitem{wang2021_graph_vul}
H.~Wang, G.~Ye, Z.~Tang, S.~H. Tan, S.~Huang, D.~Fang, Y.~Feng, L.~Bian, Z.~Wang, Combining graph-based learning with automated data collection for code vulnerability detection, IEEE Transactions on Information Forensics and Security 16 (2021) 1943--1958.
\newblock \href {https://doi.org/10.1109/TIFS.2020.3044773} {\path{doi:10.1109/TIFS.2020.3044773}}.

\bibitem{Cao2022_MVD_graph}
S.~Cao, X.~Sun, L.~Bo, R.~Wu, B.~Li, C.~Tao, \href{https://doi.org/10.1145/3510003.3510219}{Mvd: Memory-related vulnerability detection based on flow-sensitive graph neural networks}, in: Proceedings of the 44th International Conference on Software Engineering, ICSE '22, Association for Computing Machinery, New York, NY, USA, 2022, p. 1456–1468.
\newblock \href {https://doi.org/10.1145/3510003.3510219} {\path{doi:10.1145/3510003.3510219}}.
\newline\urlprefix\url{https://doi.org/10.1145/3510003.3510219}

\bibitem{wang2023_GraphSPD}
S.~Wang, X.~Wang, K.~Sun, S.~Jajodia, H.~Wang, Q.~Li, Graphspd: Graph-based security patch detection with enriched code semantics, in: 2023 IEEE Symposium on Security and Privacy (SP), 2023, pp. 2409--2426.
\newblock \href {https://doi.org/10.1109/SP46215.2023.10179479} {\path{doi:10.1109/SP46215.2023.10179479}}.

\bibitem{Wu2022_VulCNN}
Y.~Wu, D.~Zou, S.~Dou, W.~Yang, D.~Xu, H.~Jin, \href{https://doi.org/10.1145/3510003.3510229}{Vulcnn: An image-inspired scalable vulnerability detection system}, in: Proceedings of the 44th International Conference on Software Engineering, ICSE '22, Association for Computing Machinery, New York, NY, USA, 2022, p. 2365–2376.
\newblock \href {https://doi.org/10.1145/3510003.3510229} {\path{doi:10.1145/3510003.3510229}}.
\newline\urlprefix\url{https://doi.org/10.1145/3510003.3510229}

\bibitem{hin2022linevd}
D.~Hin, A.~Kan, H.~Chen, M.~A. Babar, Linevd: Statement-level vulnerability detection using graph neural networks, in: Proceedings of the 19th International Conference on Mining Software Repositories, 2022, pp. 596--607.

\bibitem{nguyen2022_ReGVD_graph}
V.-A. Nguyen, D.~Q. Nguyen, V.~Nguyen, T.~Le, Q.~H. Tran, D.~Phung, \href{https://doi.org/10.1145/3510454.3516865}{Regvd: Revisiting graph neural networks for vulnerability detection}, in: Proceedings of the ACM/IEEE 44th International Conference on Software Engineering: Companion Proceedings, ICSE '22, Association for Computing Machinery, New York, NY, USA, 2022, p. 178–182.
\newblock \href {https://doi.org/10.1145/3510454.3516865} {\path{doi:10.1145/3510454.3516865}}.
\newline\urlprefix\url{https://doi.org/10.1145/3510454.3516865}

\bibitem{guo2022hyvuldect}
W.~Guo, Y.~Fang, C.~Huang, H.~Ou, C.~Lin, Y.~Guo, Hyvuldect: A hybrid semantic vulnerability mining system based on graph neural network, Computers \& Security (2022) 102823.

\bibitem{tang2023csgvd_graph}
W.~Tang, M.~Tang, M.~Ban, Z.~Zhao, M.~Feng, Csgvd: A deep learning approach combining sequence and graph embedding for source code vulnerability detection, Journal of Systems and Software 199 (2023) 111623.

\bibitem{sard}
\href{https://samate.nist.gov/SRD/index.php}{Software assurance reference dataset}.
\newline\urlprefix\url{https://samate.nist.gov/SRD/index.php}

\bibitem{nvd}
\href{https://nvd.nist.gov/}{National vulnerability database}.
\newline\urlprefix\url{https://nvd.nist.gov/}

\bibitem{michael2005ast}
M.~Hicks, J.~S.~Foster, I.~Neamtiu, Understanding source code evolution using abstract syntax tree matching, ACM SIGSOFT Software Engineering Notes (2005).

\bibitem{wu2016google}
Y.~Wu, M.~Schuster, Z.~Chen, Q.~V. Le, Google's neural machine translation system: Bridging the gap between human and machine translation, Computer Science, Computation and Language (2016).

\bibitem{kipf2016gcn}
T.~N. Kipf, M.~Wellingd, Semi-supervised classification with graph convolutional networks, Computer Science, Machine Learning (2016).

\bibitem{lin2020neural}
G.~Lin, S.~Wen, Q.-L. Han, J.~Zhang, Y.~Xiang, Software vulnerability detection using deep neural networks: A survey, Proc. IEEE, vol. 108, issue: 10 (2020) 1825--1848.

\bibitem{daniel2018hacker}
D.~Votipka, R.~Stevens, E.~Redmiles, J.~Hu, M.~Mazurek, Hackers vs. testers: A comparison of software vulnerability discovery processes, Proc. IEEE Symp. Secur. Privacy (SP) (2018) 374--391.

\bibitem{record2018}
\href{https://www.securityweek.com/record-breaking-number\\-vulnerabilities-disclosed-2017-report}{Record-breaking number of vulnerabilities disclosed in 2017: Report.} (2018).
\newline\urlprefix\url{https://www.securityweek.com/record-breaking-number\\-vulnerabilities-disclosed-2017-report}

\bibitem{rory2019data}
R.~Coulter, Q.-L. Han, L.~Pan, J.~Zhang, Y.~Xiang, Data-driven cyber security in perspective—intelligent traffic analysis, IEEE Transactions on Cybernetics, vol. 50, issue: 7 (2019) 3081--3093.

\bibitem{ren2017rcnn}
S.~Ren, K.~He, R.~Girshick, J.~Sun, Faster r-cnn: Towards real-time object detection with region proposal networks, IEEE Transactions on Pattern Analysis and Machine Intelligence, vol. 39, issue: 6 (2017) 1137--1149.

\bibitem{yao2019textgcn}
L.~Y. et~al., Graph convolutional networks for text classification,, AAAI 33rd (2019).

\end{thebibliography}





\end{document}